\documentclass{aa}
\usepackage{graphicx}
\usepackage{natbib}
\usepackage{txfonts}
\bibpunct{(}{)}{;}{a}{}{,}
\begin{document}
%
%   \title{Putting it all together: A merger in progress in HCG31}
    \title{Is HCG31 undergoing a merger or a fly-by interaction?}

%   \subtitle{HCG31...}

   \author{Richer, M. G.
          \inst{1}
          \and
          Georgiev, L.
          \inst{2}
          \and
          Rosado, M.
          \inst{2}
          \and
          Bullejos, A.
          \inst{3,2}
          \and
          Valdez-Guti\'errez, M.
          \inst{4,5}
          \and  \\
          Dultzin-Hacyan, D.
          \inst{2}
          }

   \offprints{M. G. Richer}

   \institute{Observatorio Astron\'omico Nacional, Instituto de
              Astronom\'\i a, UNAM, P.O. Box 439027, San Diego, CA
              92143-9027\\
              \email{richer@astrosen.unam.mx}
         \and
             Instituto de Astronom\'\i a, UNAM, Apartado Postal 70-264,
             Ciudad Universitaria, 04510 M\'exico, D.F., M\'exico\\
             \email{\{georgiev, margarit, deborah\}@astroscu.unam.mx}
         \and
             Instituto de Astrof\'\i sica de Canarias, V\'\i a
             L\'actea, 38200 La Laguna, Tenerife, Spain\\
             \email{almudena@astroscu.unam.mx}
         \and
             LERMA, Observatoire de Paris, 61 Av. de
             l'Observatoire, 75014, Paris, France\\
             \email{margarita.valdez@obspm.fr}
         \and
             Instituto Nacional de Astrof\'\i sica, \'Optica y Electr\'onica,
             Apartado Postal 51, 72000 Puebla, Puebla, M\'exico\\
             \email{mago@inaoep.mx}
         }

   \date{Received <date> / Accepted <date>}

   \abstract{
   We present Fabry-Perot and multi-object spectroscopy
   of the galaxies in Hickson compact group 31 (HCG31).  Based upon
   our H$\alpha$ data cubes, galaxies A and C are a single entity,
   showing no discontinuity in their kinematics.
   Kinematically, galaxy E is probably a component of the A+C
   complex; otherwise it is a recently detached fragment. Galaxy F
   appears, both kinematically and chemically, to have formed from
   material tidally removed from the A+C complex.  Galaxies B and G
   are kinematically distinct from this complex. Galaxy Q also has
   a radial velocity compatible with group membership.  Galaxies A, B, C,
   and F have nearly identical oxygen abundances, despite spanning a
   luminosity range of 5$\,$mag.  Galaxy B's oxygen abundance is
   normal for its luminosity, while galaxy F's abundance is that
   expected
   given its origin as a tidal fragment of the A+C complex.  The
   oxygen abundances in galaxies A and C are also understandable if
   the A+C complex is a late-type spiral suffering strong gas inflow
   and star formation as a result of a tidal interaction.
   Given the kinematics of both the galaxies and the \ion{H}{i}
   gas, the oxygen abundances, and the position of galaxy G, we
   propose that an interaction of galaxy G with the A+C
   complex, rather than a merger of galaxies A and C, is a
   more complete
   explanation for the tidal features and other properties of HCG31.
   In this case, the A+C
   complex need not be a merger in progress, though this is not ruled
   out.
   \keywords{galaxies: abundances -- galaxies: evolution --
   galaxies: formation -- galaxies: individual: NGC 1741 --
   galaxies: clusters: individual: HCG31
               }
   }

   \maketitle
%
%________________________________________________________________

\section{Introduction}

The Hickson compact group of galaxies 31
\citep[HCG31;][]{1982ApJ...255..382H} has recently proven an
interesting laboratory for studying various issues related to
galaxy formation and merging. HCG31 is physically compact.  Six of
its members fall within the dimensions of a typical giant galaxy,
share a common \ion{H}{i} envelope \citep{1991AJ....101.1957W},
and span a small range in radial velocities
\citep{1990ApJ...365...86R,1992ApJ...399..353H}. It is therefore
no surprise that HCG31 shows a wide array of evidence for galaxy
interaction including: tidal tails and irregular morphology,
complex kinematics, and vigourous star formation, including
starbursts with Wolf-Rayet spectral features
\citep{1990ApJ...365...86R,1997ApJ...479..190I,1998ApJ...500..188I,
1999AJ....117.1708J,2000AJ....119.2146J}.

\citet{1990ApJ...365...86R} conclude that HCG31 will appear as a
single normal galaxy within a few orbital periods.  Here, we
differ somewhat from this conclusion.  Although we find abundant
kinematical evidence for recent gravitational interaction among
the group members, we suggest that there are other explanations
besides imminent coalescence.  Likewise, we find chemical
abundances for the different components that are unlike those of
normal dwarf irregulars, but again argue that their origins may be
easily understood.  Given the abundant evidence for gravitational
interaction, a merger may yet occur in HCG31, but we find that it
need not be currently underway.

In the following section, we present our data and their reduction.
In Sect. 3, we present our results.  In Sect. 4, we discuss these
results and their implications for the state of HCG31.  In Sect.
5, we present our interpretation of the past and future evolution
of HCG31.  In Sect. 6, we summarize our conclusions.  Throughout,
we adopt the galaxy nomenclature of \citet{1990ApJ...365...86R}
and we assume a distance of 58\,Mpc for HCG31, based upon a mean
redshift of 0.0137 \citep{1992ApJ...399..353H} and a Hubble
constant of 71\,km\,s$^{-1}$\,Mpc$^{-1}$
\citep{2000ApJ...529..786Maa}. A preliminary version of some of
our results was presented by \citet{1999IAUS..193..618R}.

\begin{figure}
\centering
\resizebox{\hsize}{!}{\includegraphics{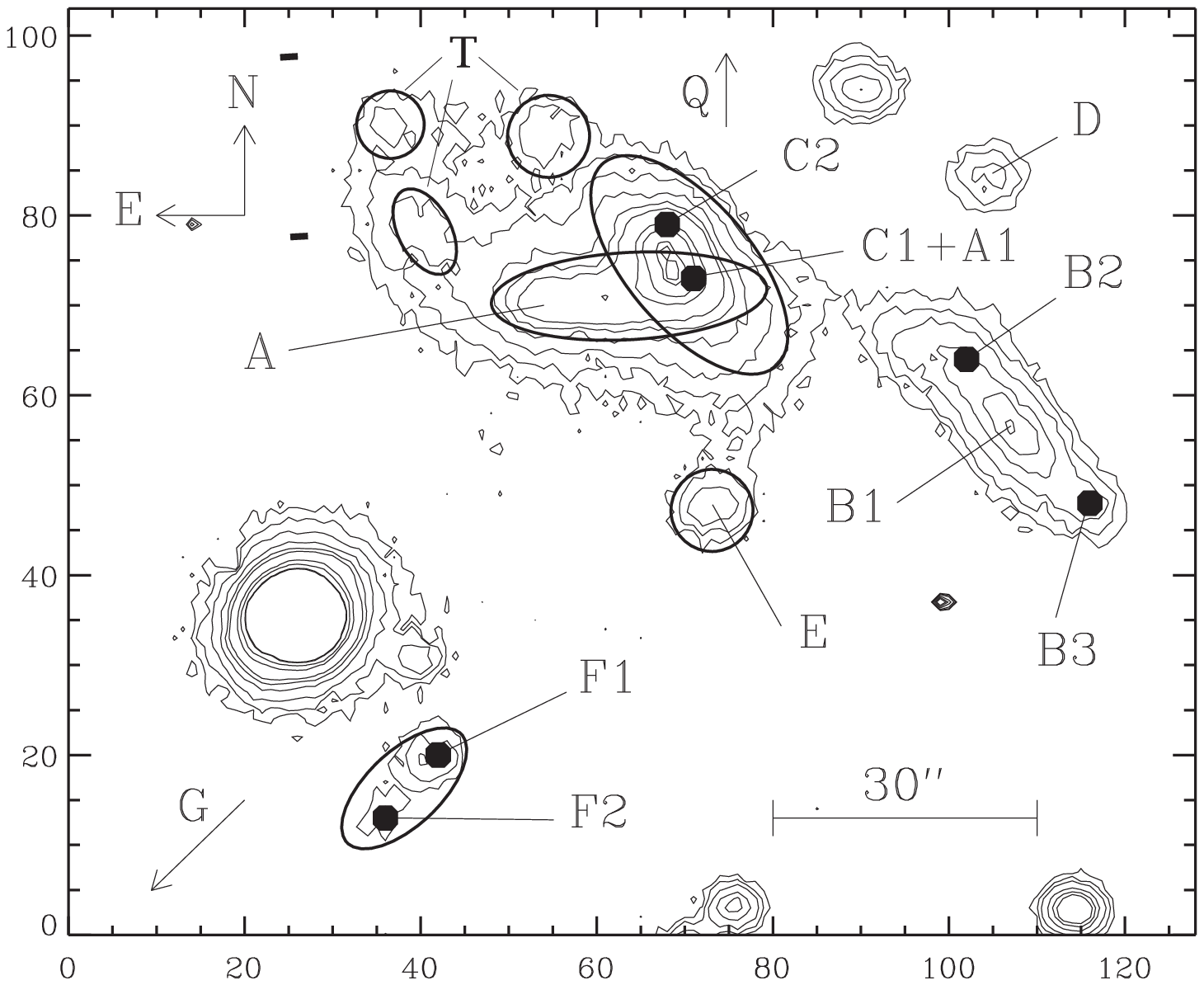}}
\caption{ The
field of HCG31 in the $V$-band.  Galaxies A, B, C, D, E, and F are
shown; galaxies G and Q are to the south-east and north,
respectively, as indicated by the arrows.  Deeper, higher-resolution
images may be found in
\citet{1990ApJ...365...86R}, \citet{1997ApJ...479..190I},
\citet{1999AJ....117.1708J}, and \citet{2000AJ....119.2146J}.
The ellipses are meant to illustrate schematically the positions
of galaxies A, C, E, F and the tidal dwarf candidates \citep[denoted
by \lq\lq T";][]{1996ApJ...462...50H}.  Objects within the
component galaxies are labelled using the galaxy component letter
and a sequence number, e.g., F1 and F2 are subcomponents of galaxy
F.  Where sub-components are labelled, the parent galaxies are not,
to reduce the clutter.
North is up and east is to the left, as indicated, and
$1\arcsec = 281$\,pc.  Unlabelled objects are foreground stars,
except that north-east of galaxy D, which is a reflection. The
filled circles indicate the locations of the apertures used for
multi-object spectroscopy.
}
\label{finder}
\end{figure}

%__________________________________________________________________

\section{Observations and reductions}

The Fabry-Perot (FP) observations of HCG31 were carried out on 1
November 1997 at the f/7.5 Cassegrain focus of the 2.1~m telescope
of the Observatorio Astron\'omico Nacional in San Pedro M\'artir,
B.C., M\'exico using the UNAM Scanning Fabry-Perot Interferometer
PUMA \citep{1995RMxAC...3..263Raa}. The scanning FP interferometer
is based upon an ET-50 Queensgate Instruments etalon with a
servo-stabilization system.  The H$\alpha$ line is observed in
interference order 330, within a free spectral range of 19.89\AA,
and sampled at 48 steps of 0.43\AA\ separation (18.9
km\,s$^{-1}$). The spectral resolution of these observations is
38.4 km~s$^{-1}$. The effective finesse obtained with this setup
was 24. Since HCG31 subtends only 2\arcmin\ and we were careful to
place it at the center of the instrument's field of view
(10\arcmin) when acquiring the data cubes, the effective finesse of
the observations is the same. In any case, PUMA does not show
important variations of the effective finesse across the field.  A
30\AA\ interference filter, centered at the wavelength of
redshifted H$\alpha$, was used to isolate the H$\alpha$ emission
line from HCG31. The detector was a 1024$\times$1024, thinned
Tektronix CCD. The image scale was 0\farcs59\,pixel$^{-1}$,
yielding a 10\arcmin\ field of view.  In order to increase the S/N
of the observations, the detector was used with a $2\times 2$
pixel binning.

We obtained two data cubes of HCG31 in H$\alpha$ with exposure
times of 48 minutes each (1 minute/channel). These data cubes were
co-added to enhance the S/N of the faint regions.  The seeing was
about 1\farcs2. The transparency conditions were rather good
and we obtained both cubes during dark time. Nevertheless, we
corrected both data cubes for transparency variations before
co-adding them. This was done using two field stars (located
outside the region shown in Fig. \ref{finder}) and we verified
that the profiles were similar.

We obtained wavelength calibration data cubes of a Ne lamp at the
beginning and end of the observations, which also serve to check
for possible equipment flexures.  Since the redshifted H$\alpha$
emission of HCG31 differs from the wavelength of the calibration
lamp, a phase shift correction was applied, which amounted to a
shift in the zero-point of the velocities of 22 km\,s$^{-1}$. The
CIGALE software package \citep{1993A&A...280..365L} was used to
apply this phase shift correction, to remove cosmic rays,
calibrate in wavelength, and construct the radial velocity cubes.
We also used some routines from the Image Reduction and Analysis
Facility (IRAF)\footnote{IRAF is distributed by the National
Optical Astronomical Observatories, which is operated by the
Associated Universities for Research in Astronomy, Inc., under
contract to the National Science Foundation.} for parts of the
data reduction.

\begin{figure}
\centering
\resizebox{\hsize}{!}{\includegraphics{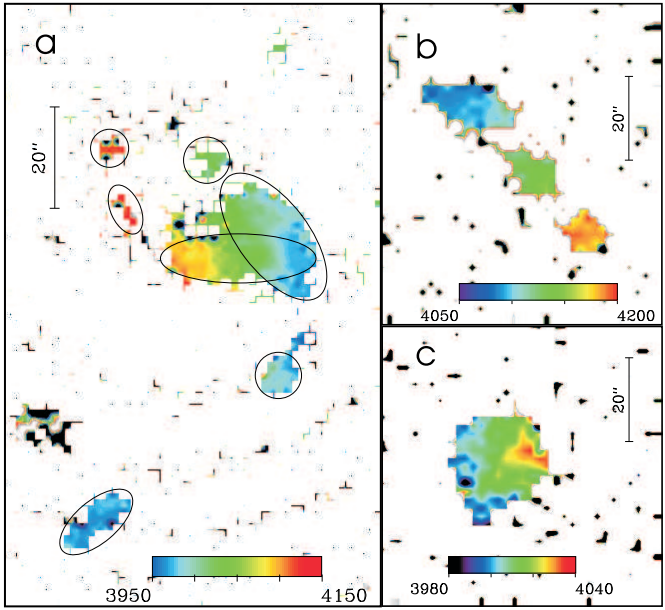}}
\caption{ The velocity map of HCG31 in H$\alpha$.  The velocities
are heliocentric velocities.  Panel (a) shows
the velocities for galaxies A, C, E, and F; panel (b) shows the
velocities for galaxy B; and panel (c) shows the velocities for
galaxy G.  In all cases, the colour bars show the total range of
radial velocities.  The ellipses are again meant to illustrate
the positions of galaxies A, C, E, F, and the tidal candidates,
as in Fig. \ref{finder}.  In general, both our radial velocities and
velocity profiles are in good agreement with those of
\citet{1990ApJ...365...86R}.  The continuity of the kinematics for
galaxies A, C, and E is particularly striking, and probably
indicates that these three components are  a single entity.
Galaxy Q was not detected in our Fabry-Perot data cubes, while
the H$\alpha$ line from galaxy D is shifted outside our filter by
this galaxy's much higher radial velocity
\citep{1992ApJ...399..353H}.  In all panels, north is up, east is
to the left, and $1\arcsec = 281$\,pc.
}
\label{radvel}
\end{figure}

\begin{figure}
\centering \resizebox{\hsize}{!}{\includegraphics{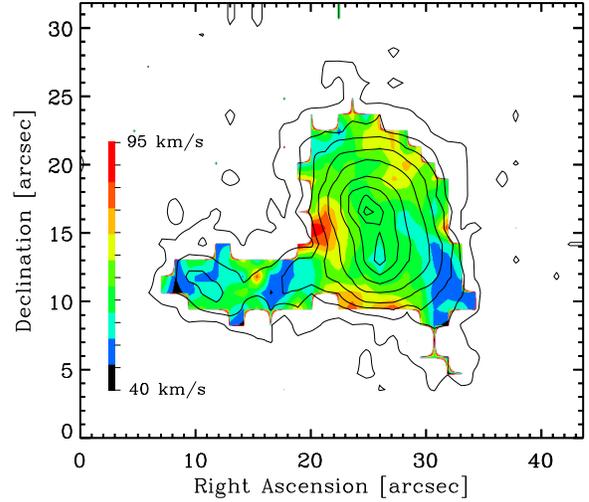}}
\caption{ The velocity dispersion map of galaxies A and C in
H$\alpha$. The colour scale shows the range in velocity dispersion
values. Here, we characterize the velocity dispersion as the
standard deviation of the gaussian fit to the velocity profile.
Only galaxies A and C showed significant internal variations in
velocity dispersion, so they are the only ones shown here.  The
contours show the total H$\alpha$ flux from the Fabry-Perot
observations.  As usual, north is up, east is to the left, and
$1\arcsec = 281$\,pc. } \label{veldisp}
\end{figure}

The multi-object spectroscopy of the galaxies in \object{HCG31}
was obtained with the 2.1m telescope of the Observatorio
Astrof\'\i sico Guillermo Haro in Cananea, M\'exico on 10 January
1999.  The LFOSC spectrograph was used, which is a transmission
spectrograph employing a grism as the dispersing element
\citep{1997A&AS..123..103Zaa}.  The spectrograph's dispersion was
approximately 5.5\,\AA/pix and the spectra typically spanned from
4000\AA\ to 6600\AA, though the exact spectral range varied
according to the object's position within the spectrograph's field
of view.  The detector was a $576\times 384$ EEV CCD.  Since the
spectrograph has very poor response in the blue, no order-sorting
filter was used. The objects were selected for spectroscopy using
focal plane masks made from previously-acquired images.  The only
restriction for object selection is that they may not be aligned
in declination since the dispersion axis is oriented east-west.
Pairs of holes were cut for object and sky at the same right
ascension so as to ensure identical spectral coverage.  The holes
were 3\arcsec\ in diameter.

The objects selected for spectroscopy are shown in Fig.
\ref{finder}, with the exception of galaxy Q
\citep{1990ApJ...365...86R}, which is to the north of the field.
The total integration time was 1.5 hours.

The standard stars \object{HD 19445}, \object{HD 74721}, and
\object{HD 109995} were observed for flux calibration.  The
standard stars were observed through masks cut for various object
fields.

The multi-object spectroscopy was reduced using the IRAF software
package, specifically the noao.imred.specred package. The bias images were
first combined and the result subtracted from all of the images.
Pixel-to-pixel variations were then removed using spectra of the
internal lamp.  Subsequently, the sky was subtracted from each
object.  Spectra of the Ne-Ar lamp were used for wavelength
calibration.  Finally, the flux calibration was made using the
standard star observations.

%__________________________________________________________________

\section{Results}

\begin{figure}
\centering \resizebox{\hsize}{!}{\includegraphics{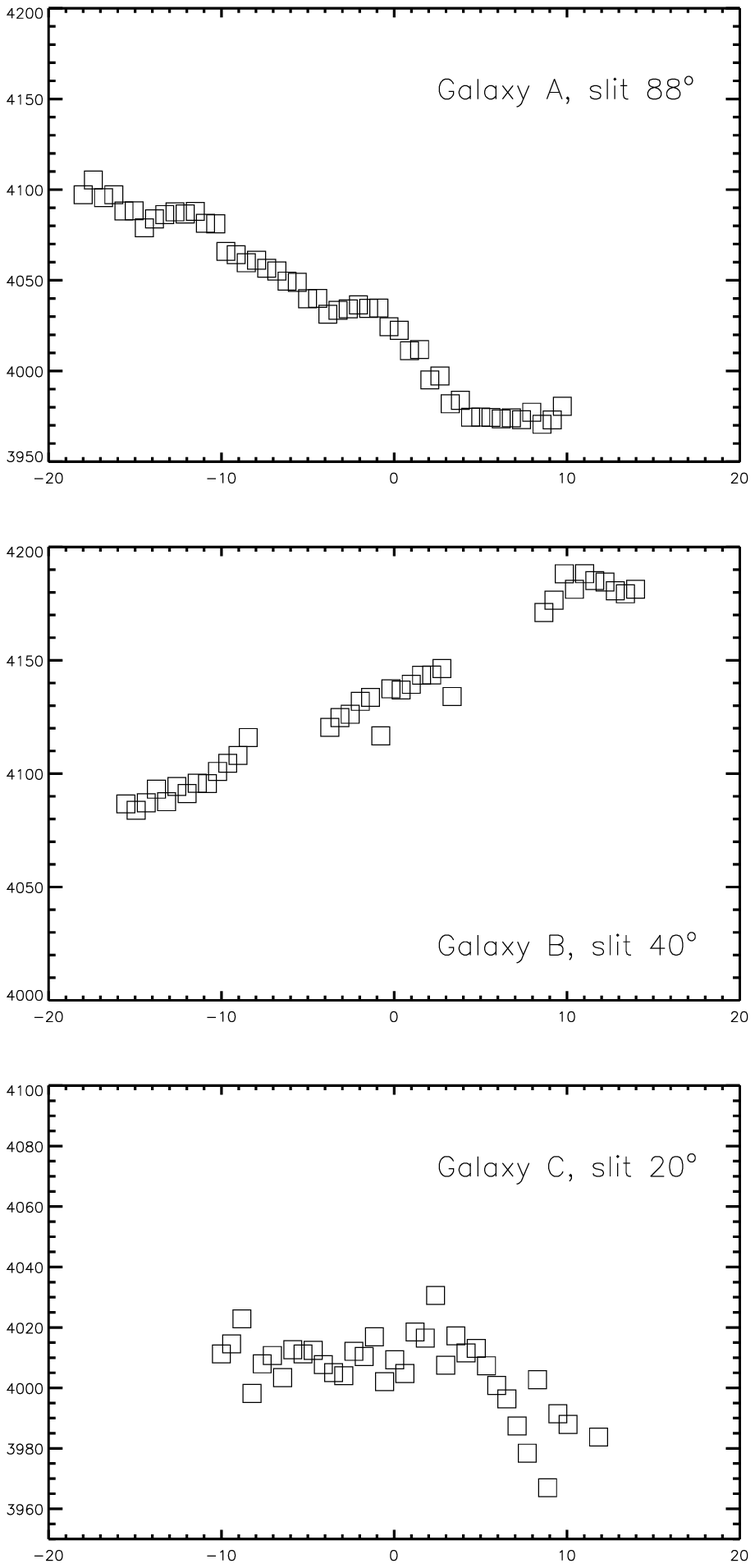}}
\caption{ This figure presents velocity profiles of galaxies A, B,
and C extracted at the same positions and position angles as
\citet{1990ApJ...365...86R}.  Each panel indicates the position
angle of the simulated slit.  The agreement is very good for both
the velocities and the velocity profiles, any discrepancies being
of order 10\,km/s (compare with their Figs.
2 and 3).  We are not as sensitive to low surface brightness
emission, so our profiles have a smaller spatial extent than
those of \citet{1990ApJ...365...86R}.
}
\label{velprof}
\end{figure}

Figs. \ref{radvel} and \ref{veldisp} present the radial velocity
and velocity dispersion maps resulting from our Fabry-Perot
spectroscopy at H$\alpha$. To compute these maps, we computed the
average intensity and its standard deviation over all channels at
each spatial pixel. Then, we determined the median value of this
standard deviation over all spatial pixels in the image and
adopted this as representative of the noise in the sky throughout
the image. Finally, at each spatial pixel, the maximum of the
intensities in all channels was compared to this noise value. For
those spatial pixels where the maximum intensity in some channel
exceeded five times the noise value, a gaussian was fit to the
velocity profile. The peak position and the standard deviation of
this gaussian fit were adopted as the radial velocity and velocity
dispersion, respectively, at that spatial position.

In Fig. \ref{radvel}, our radial velocities and velocity profiles
are in good agreement with those found by
\citet{1990ApJ...365...86R}.  In Fig. \ref{velprof}, we present
several simulated profiles for the positions and position angles
corresponding to Figs. 2 and 3 of \citet{1990ApJ...365...86R}.
Galaxy Q was not detected in our data cubes, while galaxy D's much
higher radial velocity shifted its H$\alpha$ line outside our
filter.

In Fig. \ref{veldisp}, we present the velocity dispersions in
galaxies A and C.  We present a map of the velocity dispersion for
galaxies A and C only, since no other galaxies showed significant
internal variations in velocity dispersion. The velocity
dispersions in galaxies B, E, and F are approximately
50\,km\,s$^{-1}$, similar to our instrumental resolution, while
the velocity dispersion in galaxy G is slightly higher.

\begin{table*}
\caption[]{Raw line intensities$^\mathrm{a}$ for the \ion{H}{ii}
regions in HCG31 measured from the multi-object spectroscopy}
\label{inttable}
\[
\begin{tabular}{lccccccc}
\hline \noalign{\smallskip} $\lambda$                     & A1+C1
&    C2           &   B2           &    B3           &    F1 &
F2           &   Q           \\
\noalign{\smallskip}\hline\noalign{\smallskip} H$\epsilon$\hfill
3970 & $             $ & $12.51\pm 0.80$ & $            $ & $ $ &
$             $ & $             $ & $           $ \\
H$\delta$\hfill          4101 & $10.44\pm 0.99$ & $18.42\pm 0.86$
& $            $ & $             $ & $ 10.3\pm  2.7$ & $ $ & $
$ \\ H$\gamma$\hfill          4340 & $ 29.2\pm 1.3$ & $ 38.4\pm
1.1$ & $            $ & $ 33.6\pm  3.3$ & $ 31.1\pm  3.1$ & $
31.9\pm  5.3$ & $           $ \\ {[}\ion{O}{iii}{]}\hfill 4363 & $
2.20\pm 0.93$ & $ 1.21\pm 0.61$ & $            $ & $             $
& $  5.1\pm  2.5$ & $ $ & $           $ \\ \ion{He}{i}\hfill
4472 & $ 1.14\pm 0.97$ & $ 2.70\pm 0.66$ & $            $ & $
$ & $ $ & $             $ & $           $ \\ H$\beta$\hfill 4861 &
$100.0\pm  2.8$ & $100.0\pm  2.9$ & $ 100\pm   16$ & $100.0\pm
3.5$ & $100.0\pm  2.6$ & $100.0\pm  4.5$ & $100\pm 32$ \\
{[}\ion{O}{iii}{]}\hfill 4959 & $ 71.6\pm  3.2$ & $ 90.0\pm 3.9$ &
$ 135\pm   19$ & $117.2\pm  4.6$ & $171.8\pm  6.3$ & $142.9\pm
6.4$ & $           $ \\ {[}\ion{O}{iii}{]}\hfill 5007 & $218.1\pm
5.4$ & $273.5\pm  6.9$ & $ 406\pm   47$ & $361.3\pm 9.8$ & $
521\pm   12$ & $  429\pm   15$ & $157\pm   43$ \\
\ion{He}{i}\hfill        5876 & $15.68\pm 0.43$ & $11.67\pm 0.42$
& $            $ & $ 12.0\pm  1.3$ & $13.83\pm 0.91$ & $ 13.8\pm
1.6$ & $           $ \\ {[}\ion{O}{i}{]}\hfill   6300 & $ 8.77\pm
0.42$ & $ 4.77\pm 0.39$ & $            $ & $ 10.7\pm  1.3$ & $
8.06\pm 0.93$ & $             $ & $           $ \\
{[}\ion{O}{i}{]}\hfill   6364 & $ 2.39\pm 0.28$ & $ 1.30\pm 0.13$
& $            $ & $ 1.51\pm 0.92$ & $ 2.42\pm 0.67$ & $ $ & $
$ \\ H$\alpha$\hfill          6563 & $  473\pm 10$ & $326.4\pm
7.2$ & $ 625\pm   71$ & $             $ & $367.4\pm  7.1$ & $
385\pm   13$ & $470\pm  109$ \\ {[}\ion{N}{ii}{]}\hfill  6583 & $
74.6\pm  3.9$ & $ 34.1\pm  2.4$ & $  57\pm   12$ & $             $
& $ 11.8\pm  1.7$ & $ 12.9\pm 2.5$ & $ 51\pm   20$ \\
\ion{He}{i}\hfill        6678 & $  5.4\pm 1.0$ & $ 3.55\pm 0.34$ &
$            $ & $             $ & $ 5.23\pm 0.88$ & $
$ & $           $ \\ {[}\ion{S}{ii}{]}\hfill  6716 & $ 47.8\pm
2.1$ & $25.01\pm 0.98$ & $            $ & $             $ & $
24.8\pm  1.9$ & $ 23.6\pm 2.5$ & $146\pm   37$ \\
{[}\ion{S}{ii}{]}\hfill  6731 & $ 38.7\pm 1.8$ & $17.54\pm 0.77$ &
$            $ & $             $ & $ 14.6\pm  1.7$ & $  9.3\pm
2.4$ & $125\pm   33$ \\ V$_r ({\mathrm H}\alpha$)\hfill (km/s) &
3879   & 3947            & 4068 &                 & 3843
& 3831            & 4148
\\ $E(B-V)$ \hfill (mag)         & $ 0.42\pm 0.02$ & $ 0.10\pm
0.02$ & $0.42\pm 0.10$ & $ 0.68\pm 0.44$ & $ 0.21\pm 0.02$ & $
0.26\pm 0.03$ &0$.11\pm 0.22$ \\ $T_e([\ion{O}{iii}])$\hfill (K)&
$12500\pm 3200$ & $ 9100\pm 2000$ & $            $ & $ $ &
$11900\pm 3200$ & $             $ & $           $ \\
$12+\log({\mathrm O}^{++}/{\mathrm H})$ & $7.56\pm 0.34$ &
$8.12\pm 0.39$ &        &                 & $8.02\pm  0.38$ & &
\\ $12+\log({\mathrm O}/{\mathrm H})_{Te}$ & $8.1\pm 0.2  $ &
$8.3\pm   0.2$ &        &                 & $8.1\pm    0.2$ &
&               \\ \noalign{\smallskip}\hline
\end{tabular}
\]
\begin{list}{}{}
\item[$^{\mathrm a}$] The line intensities are measured relative
to H$\beta$ and are not corrected for interstellar reddening.
\end{list}
\end{table*}

Table \ref{inttable} lists the raw line intensities relative to
H$\beta$ and the raw radial velocities we measured in the
\ion{H}{ii} regions in the different components of HCG31.  The
uncertainties quoted for the line intensities are 1$\sigma$
uncertainties and include the uncertainty in the fits to both the
line and the reference line as well as the uncertainty in the
continuum placement for both lines.  The line intensities are not
corrected for reddening. The names coincide with those indicated
in Fig. \ref{finder}. Table \ref{inttable} also includes the
reddening derived from the Balmer lines along with the abundance
of doubly ionized oxygen and an estimate of the total oxygen
abundance for those cases when we could derive an electron
temperature.

Before computing the reddenings listed in Table \ref{inttable}, we
corrected the Balmer lines for underlying absorption according to
\begin{equation}
$$F_c(\lambda) = F_o(\lambda) \times
\left(1+\frac{W_\mathrm{abs}}{W_\lambda}\right) /
\left(1+\frac{W_\mathrm{abs}}{W_{\mathrm H \beta}}\right)
\label{eqnwidth}$$,
\end{equation}
where $F_o(\lambda)$ and $F_c(\lambda)$ are the observed and
corrected fluxes and $W_\mathrm{abs}$, $W_\lambda$, and
$W_{\mathrm H \beta}$ are the equivalent widths of the underlying
stellar absorption, of the emission line in question, and of
H$\beta$, respectively.  We used $W_\mathrm{abs} = 1.9$\AA\
\citep{1985ApJS...57....1M} since our spectra had neither the
resolution nor the signal-to-noise required to measure this
directly. These corrected line intensities were then used to
compute the reddening via
\begin{equation}
$$\log \frac{F(\lambda)}{F(\mathrm H\beta)} = \log
\frac{I(\lambda)}{I(\mathrm H\beta)} - 0.4
E(B-V)\left(A_1(\lambda) - A_1(\mathrm H\beta)\right)
\label{eqnredd}$$,
\end{equation}
where $F(\lambda)$ and $I(\lambda)$ are the observed and
theoretical emission-line fluxes at wavelength $\lambda$, $E(B-V)$
is the reddening, and $A_1(\lambda) = A(\lambda)/E(B-V)$.  Values
of $A_1(\lambda)$ were taken from the \citet{1977AJ.....82..337S}
reddening law. We used the ${\mathrm H}\alpha/{\mathrm H}\beta$
ratio to determine the reddening, except when H$\alpha$ fell
outside our wavelength range, in which case the ${\mathrm
H}\gamma/{\mathrm H}\beta$ ratio was used.  In all cases, we
adopted theoretical ratios of $I({\mathrm H}\alpha)/I({\mathrm
H}\beta) = 2.86$ and $I({\mathrm H}\gamma)/I({\mathrm H}\beta) =
0.468$, appropriate for electron temperatures and densities of
$10^4$\,K and 100\,cm$^{-3}$, respectively
\citep{1989agna.book.....O}.

The radial velocities given in Table \ref{inttable} are the raw
radial velocities measured for the H$\alpha$ line from the
multi-object spectra, and are meant primarily to show that galaxy
Q has a radial velocity consistent with group membership, and
similar to the \ion{H}{i} velocity found by
\citet{1991AJ....101.1957W}. Galaxy Q is the faintest of the
components we observed, which may explain why it is so faint in
the H$\alpha$ image presented by \citet{1990ApJ...365...86R} and
absent from our H$\alpha$ data cubes.  We cannot restrict galaxy
Q's morphological type since the $V$-band and H$\alpha$ images we
obtained to identify \ion{H}{ii} regions are of low spatial
resolution and signal-to-noise and were never calibrated.

The abundance of doubly ionized oxygen is computed according to
the prescription given in \citet{1993ApJ...415..240R} using the
reddening-corrected line intensities and assuming an electron
density of 100\,cm$^{-3}$. The total oxygen abundances are
estimated from Fig. 5 of \citet{1999ApJ...514..544K} in those
cases when the electron temperature was measured. Generally, these
total oxygen abundances are in good agreement with extant data
(see Table \ref{lohtable}).

\begin{table}
\caption[]{Luminosities and oxygen abundances} \label{lohtable}
\[
\begin{tabular}{lccll}
\hline \noalign{\smallskip} Galaxy & $B$ &
$12+\log({\mathrm O}/{\mathrm H})$ & method$^{\mathrm b}$ & ref
\\ \noalign{\smallskip}\hline \noalign{\smallskip}
A      & $15.62\pm 0.5$ & $8.04\pm 0.06$ & A   & 1,3\\
B      & $15.35\pm 0.2$ & $8.34\pm 0.20$ & B   & 1,4\\
C      & $13.44\pm 0.5$ & $8.3\pm 0.2$   & A,C &1,5,6\\
F      & $18.2\pm 1.0\,^{\mathrm a}$ & $8.1\pm 0.2$ & C &2,6\\
\noalign{\smallskip}\hline
\end{tabular}
\]
\begin{list}{}{}
\item[$^{\mathrm a}$] The luminosity of component F is assumed equal to that of
component E.
\item[$^{\mathrm b}$] The oxygen abundances are obtained by the
following methods: (A) measurements of [\ion{O}{ii}] and
[\ion{O}{iii}] lines and the electron temperature; (B)
measurements of [\ion{O}{ii}] and [\ion{O}{iii}] lines only; (C)
measurement of the [\ion{O}{iii}] lines and the electron
temperature
\end{list}
References: (1) \citealt{1988ApJ...331...64H}; (2)
\citealt{1997ApJ...479..190I}; (3) \citealt{1998ApJ...500..188I};
(4) \citealt{1990ApJ...365...86R}; (5)
\citealt{1992ApJ...401..543V}; (6) Table \ref{inttable};
\end{table}

\begin{figure}
\resizebox{\hsize}{!}{\includegraphics{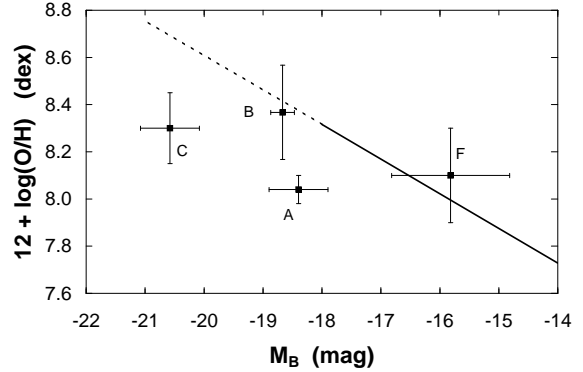}} \caption{
The relation between luminosity and metallicity for the galaxies
in HCG31.  The components are labelled. The solid line is the
relation for dwarf irregulars found by
\citet{1995ApJ...445..642R}, while the dashed line is an
extrapolation of that relation.  The galaxies in HCG31 are clearly
discrepant, with components A and C much too luminous for their
abundances. } \label{metlum}
\end{figure}

In Table \ref{lohtable}, we list the luminosities and oxygen
abundances for the component galaxies with measured oxygen
abundances. For each galaxy, we adopted what appeared to be the
best abundance determination available, indicated in the footnotes
to Table \ref{lohtable}, using the precision quoted in the
original papers as our guide. For galaxies A, B, and C, we adopted
the \citet{1988ApJ...331...64H} $B$-band magnitudes at the
24.5$\,\mathrm {mag}/\sq\arcsec$ isophote since these seem to be
in better agreement with the photometry of
\citet{1997ApJ...479..190I}.  No photometry exists for galaxy F.
Here, we assume it has the same luminosity as galaxy E
\citep{1997ApJ...479..190I}, which, though it is a rough
approximation, is sufficient for our purposes.  In Table
\ref{lohtable}, we adopted the uncertainties in the luminosities
of galaxies B and C from \citet{1988ApJ...331...64H}. For galaxy
A, we set the uncertainty in luminosity equal to that for galaxy C
due to the difficulty of disentangling the two galaxies.  We
arbitrarily estimate an uncertainty of 1.0 mag in the luminosity
of galaxy F. In Fig. \ref{metlum}, we plot the oxygen abundance as
a function of the galaxy luminosity.  We converted from apparent
to absolute magnitudes using an apparent distance modulus of
34.02\,mag, which accounts for our adopted distance and an
extinction of $A_B = 0.21$\,mag, based upon $E(B-V) = 0.055$\,mag
\citep{1998ApJ...500..525S} and the \citet{1999PASP..111...63F}
reddening law ($R_V = 3.1$ and $E(B-V) = 1$\,mag).  We also show
the relation between oxygen abundance and luminosity for dwarf
irregulars for comparison \citep{1995ApJ...445..642R}.

%
%---------------------------------------------------------------

\section{The state of HCG31 today}

Our most surprising result is that galaxies A and C appear to be a
single kinematical entity.  In Fig. \ref{radvel}, their kinematics
show no discontinuity.  We interpret this as evidence that
galaxies A and C are a single entity at present.
\citet{2000sgg..conf...60M} and \citet{2000dgeu.conf..369A} find
similar results from similar kinematical data.

Another surprise is that galaxy E's position and velocity are
compatible with it also being an integral part of the A+C complex.
If galaxy E is in fact a separate entity, it is just now
separating from the A+C complex.  In any case, its kinematics
confirm that the extension of the A+C complex towards galaxies E
and F is a tidal tail.

Like galaxy E, the tidal dwarfs to the north-east of galaxies A
and C identified by \citet{1996ApJ...462...50H} also have
kinematics consistent with their being integral components of the
A+C complex. Furthermore, given the red continuum image presented
by \citet{2000AJ....119.2146J}, it is not evident that at least
the nearer of these supposed tidal dwarfs are separate entities.

Meanwhile, galaxy F is an excellent example of a tidal fragment.
Considering its position and radial velocity (Figs. \ref{finder}
and \ref{radvel}), it seems clear that galaxy F is a tidal
fragment of the A+C complex that detached some time ago.
Remarkably, galaxy F has the same oxygen abundance as galaxies A
and C, despite being more than 5\,mag fainter than galaxy C (cf.
Table \ref{lohtable} and Fig. \ref{metlum}).  In addition, since
an underlying stellar population has yet to be detected in galaxy
F \citep{2000AJ....119.2146J}, it cannot be responsible for its
own chemical enrichment.  All of these observations argue that
galaxy F is a tidal fragment of the A+C complex.
%By extension,
%all of these observations also provide evidence of past tidal
%interaction between galaxies A and C.

Galaxy B exhibits solid-body rotation.  Given its broadband
morphology \citep{1990ApJ...365...86R,1997ApJ...479..190I}, it is
likely a dwarf spiral or irregular seen nearly edge-on.
Kinematically, galaxy B is distinct from the A+C complex and is
counter-rotating with respect to it.
%This discounts the
%suggestion of \citet{2000AJ....120..621P} that the three central
%galaxies of HCG31 may be single irregular galaxy, but instead
%shows that galaxy B is an independent galaxy.

Galaxy G appears to be a small disk galaxy seen nearly face-on
since the H$\alpha$ velocity has only a very small gradient across
its disk.

The velocity dispersion map (Fig. \ref{veldisp}) of the ionized
gas is more difficult to interpret.  The highest velocity
dispersions occur along a ridge between the main bodies of
galaxies A and C. Whether this higher velocity dispersion is due
to a more complicated gas distribution or kinematics is
indeterminate from our observations.

It is evident from Fig. \ref{metlum} that the oxygen abundances in
the galaxies in HCG31 are unusual.  Although they span 5\,mag in
luminosity, all of the galaxies have nearly identical oxygen
abundances.  This conclusion is independent of the uncertainty in
separating the luminosities of galaxies A and C, since, were we to
consider their combined luminosity and either of the oxygen
abundances observed (Table \ref{lohtable}), the resulting position
would remain highly discrepant compared to the relation for normal
dwarf irregular galaxies.  However, galaxy C's luminosity is much
greater than that of most dwarfs and more like that of a typical
late-type spiral.  If the A+C complex were a late-type spiral, its
oxygen abundance may be more easily understood.  Given the strong
inflows and consequent high star formation rate expected due to
tidal interactions
\citep[e.g.,][]{1996ApJ...464..641M,1996ApJ...471..115B}, the mass
of gas expected in the outer disks of late-type spirals, and the
low oxygen abundances observed in these regions
\citep{1998AJ....116..673F,1998AJ....116.2805V}, the unusual
positions of galaxies A and C in Fig. \ref{metlum} are
understandable as the result of dilution due to strong gas inflow
and the transient high luminosity due to a starburst. At any rate,
galaxy C's position is compatible with the extreme low end of the
abundance range for late-type spirals \citep{1997ApJ...489...63G}.
Given galaxy F's origin within the A+C complex, it should
naturally have the same oxygen abundance as the A+C complex.

Galaxy B, meanwhile, appears more or less normal.  It has a
$B$-band luminosity at the bright end of the range found for dwarf
irregulars, about 0.8\,mag brighter than the LMC's, and an oxygen
abundance similar to that of the LMC.  Its H$\alpha$ luminosity,
$1.0\times 10^{41}\,\mathrm{erg}\,\mathrm s^{-1}$ \citep[
corrected to our adopted distance]{1997ApJ...479..190I}, implies a
global star formation rate of $0.8\,M_\odot \,\mathrm{yr}^{-1}$
\citep{1998ARA&A..36..189K}, which is not unusual for a late-type
galaxy \citep{1983ApJ...272...54K}.  However, its colours are
significantly bluer than those of typical dwarf irregulars or
late-type spirals
\citep{1997ApJ...479..190I,1981AJ.....86.1429D,1994ARA&A..32..115R},
being a near-perfect match to those of a single stellar population
of metallicity $Z=0.008$ and an age of 40\,Myr
\citep{1994ApJS...94...63B}.  Thus, galaxy B is currently forming
stars more actively than it has formed them on average in the
past, perhaps as a result of the group environment.  In spite of
this, galaxy B's recent history of star formation appears to be
distinct from that of the A+C complex
\citep{1997ApJ...479..190I,2000AJ....119.2146J}.

\section{The past and future}

The kinematical evidence argues strongly that the A+C complex has
undergone very significant, recent, tidal interaction, but that it
is now a single entity.  In addition, the kinematical properties
of galaxies E, F, and the tidal candidates identified by
\citet{1996ApJ...462...50H} all provide evidence of their origin
within the A+C complex.  What, then, are the possible origins of
these tidal interactions, and might these shed some light on the
nature of HCG31?  A related issue is that of the nature of the
original components of HCG31.

The global \ion{H}{i} mass and blue luminosity of HCG31 provide
some clues regarding the original composition of HCG31.  The
\ion{H}{i} mass of HCG31 is strikingly high, $2.5\times
10^{10}\,M_\odot$ adjusted to our adopted distance
\citep{1991AJ....101.1957W}, which would be typical of a massive
Sb or Sc spiral \citep{1994ARA&A..32..115R}.  The oxygen
abundances, however, are incompatible with those of such massive
galaxies save those found in their extreme outer disks
\citep{1997ApJ...489...63G,1998AJ....116..673F,1998AJ....116.2805V}.
The blue luminosities of galaxies A+C, B, and G imply
morphological types Scd, Sm, and Sd \citep{1994ARA&A..32..115R},
though these are all undoubtedly affected by their significant
recent star formation. The total \ion{H}{i} mass is also
consistent with the sum of three late-type spirals \citep[or one
such spiral and two bright irregulars;][]{1994ARA&A..32..115R}. As
noted above, the oxygen abundances of A+C and B are compatible
with such progenitors, though that for A+C would be at the low end
of the range expected
\citep{1995ApJ...445..642R,1997ApJ...489...63G}. Finally, the
global $L_B/M(\ion{H}{i})$ for the entire group is approximately
unity, a value typical of dwarf irregulars
\citep{1994ARA&A..32..115R}.  Despite the uncertainty due to the
recent star formation, all of the available evidence points toward
the group being initially composed of late-type spirals and
irregulars.

The kinematical evolution of HCG31 is a more ambiguous puzzle.
Based upon the galaxy morphologies and less complete kinematical
data than we present here, an on-going merger between galaxies A
and C has been invoked to explain the tidal features observed
\citep[e.g.,][]{1990ApJ...365...86R}. However, this explanation
leaves a number of issues unresolved, the two most important of
which are the coincidence of galaxy G with a peak in the
\ion{H}{i} distribution at the end of a tidal tail and the very
similar recent star formation histories in galaxies A, C, E, F,
and G
\citep{1991AJ....101.1957W,1997ApJ...479..190I,1999AJ....117.1708J}.
The continuity of the kinematics of the A+C complex is an
additional important complication since it implies that the
interaction would be in an advanced, settled stage, in contrast to
numerical simulations that indicate that most of the star
formation ought to occur at the onset of the interaction
\citep{1996ApJ...471..115B,1996ApJ...464..641M}.  Likewise, the
blue colours of the star clusters in galaxies A and C imply that
any merger should be in an initial phase
\citep{1999AJ....117.1708J}.

In view of the continuous kinematics of the A+C complex \citep[see
also][]{2000dgeu.conf..369A}, we propose that an encounter between
it and galaxy G is a more likely origin for the tidal interaction
that is so evident. This encounter scenario naturally accounts for
the position of galaxy G, embedded in the south-east tidal tail of
the \ion{H}{i} distribution \citep{1991AJ....101.1957W}.  An
encounter between galaxy G and the A+C complex also easily
explains the very similar histories of the most recent star
formation in galaxies A+C, E, F, and G noted by
\citet{1997ApJ...479..190I} and \citet{2000AJ....119.2146J},
despite their physical separation of up to 40\,kpc, since such an
encounter would provoke strong gas inflow and presumably vigourous
star formation in galaxy G, the A+C complex, and in the tidal
material
\citep{1992ApJ...400..153M,1996ApJ...464..641M,1996ApJ...471..115B}.
Such strong gas inflow would also explain the low oxygen abundance
measured in the A+C complex as a result of dilution by metal-poor
gas. The coordinated star formation, the global morphology and
kinematics of the \ion{H}{i} gas \citep{1991AJ....101.1957W}, and
the positions of galaxies A+C, E, F, and G are all very
reminiscent of the fly-by and merger simulations of
\citet{1992ApJ...400..153M}, \citet{1996ApJ...464..641M}, and
\citet{1996ApJ...471..115B}. Finally,
\citet{2002ApJ...575..747Oaa} conclude that the burst of star
formation in the A+C complex may be evolving to a post-starburst
phase, based upon ISO data, a result compatible with the encounter
scenario if galaxy G is now moving away from the A+C complex.
Therefore, unless galaxy G occupies its position purely by chance,
we propose that an encounter between it and the A+C complex
accounts for more of the properties of HCG31 in a more natural
fashion than does a merger of galaxies A and C.

An important uncertainty in the encounter scenario is the
transverse velocity implied between galaxy G and the A+C complex.
Given their very similar radial velocities, the bulk of their
relative motion must be in the plane of the sky \citep[Fig.
\ref{radvel};][]{1990ApJ...365...86R,1992ApJ...399..353H}.  At a
distance of 58\,Mpc, the 2\farcm4 separation of galaxy G and the
A+C complex corresponds to a linear distance of 40\,kpc. If the
ages of the most recent star formation, $< 10^8$\,years
\citep{1997ApJ...479..190I,1999AJ....117.1708J}, are taken as the
time elapsed since the putative encounter, the separation implies
a transverse velocity in the plane of the sky exceeding
400\,km\,s$^{-1}$. This velocity greatly exceeds the internal
velocity dispersions we measure in these galaxies (Fig.
\ref{veldisp}; Sect. 3), and it is not obvious whether this
encounter velocity should provoke such a strong interaction
\citep[e.g.,][]{1987gady.book.....B}. The time scale could be
lengthened, and the transverse velocity lowered, if there is a
delay between closest approach and the initiation of star
formation, as some models indicate
\citep[e.g.,][]{1992ApJ...400..153M,1996ApJ...464..641M}.  On the
other hand, neither the velocity nor the spatial extension between
galaxy G and the A+C complex are unusual. HCG26 is very similar
\citep{1995grga.conf...77W}, while HCG16 has a similar morphology
on a larger scale \citep{2001A&A...377..812Vaa}. NGC 7318b in
Stephan's Quintet (HCG92) has been stripped of its \ion{H}{i}
despite a velocity of order 1000\,km\,s$^{-1}$ relative to the
other group members \citep{2001AJ....122.2993Saa}. Likewise, UGC
7636 in the Virgo cluster has had its interstellar medium stripped
by NGC 4472 though their relative velocities differ by about
700\,km\,s$^{-1}$
\citep{1992ApJ...400L..55P,1994AJ....108..844M,2000ApJ...530L..17L}.
Therefore, although the response of the \ion{H}{i} in fly-by
encounters appears to be relatively unexplored theoretically,
observational examples appear reasonably common, lending credence
to the idea that an interaction between galaxies G and A+C could
be the origin of the tidal debris seen in HCG31.

If the gravitational interaction between galaxy G and the A+C
complex is the origin of the tidal debris in HCG31, it is then
also possible that the A+C complex has always been a single entity
and need not represent the fusion of two galaxies, partially
confirming the hypothesis of \citet{2000AJ....120..621P}. The
simulations of fly-by and merger interactions indicate that very
strong gas inflows may be induced
\citep{1992ApJ...400..153M,1996ApJ...464..641M,1996ApJ...471..115B}
and such inflows are indeed observed in HCG31
\citep{1991AJ....101.1957W}, though the \ion{H}{i} kinematics are
slightly compromised since their spatial resolution is
insufficient to clearly separate the gas belonging to galaxy B
from that belonging to the A+C complex. The distribution of the
star formation observed in the A+C complex could simply be a
typical result of the gas inflows predicted by the numerical
simulations.  Likewise, the low oxygen abundances could easily
result from dilution due to the inflow of metal-poor material from
the outer disk.

What might be the eventual outcome of HCG31?  If an encounter
between galaxy G and the A+C complex is the origin of the tidal
interactions, and if their orbits are unbound, the A+C complex
will likely evolve similar to the simulations of ring galaxies
\citep[e.g.,][]{1997ApJS..113..269S}, and the A+C complex should
remain as a disk galaxy.  If, on the contrary, their orbits are
bound, a merger that will likely also consume galaxies E, F, and
the tidal dwarfs postulated by \citet{1996ApJ...462...50H} is the
probable result.  Finally, if the interaction between galaxies A
and C is the origin of the tidal debris in HCG31, as has been
usually supposed
\citep[e.g.,][]{1990ApJ...365...86R,1999AJ....117.1708J,2000sgg..conf...60M},
a merger already appears to be well underway.

In either case of a merger, we expect that a low mass elliptical
would be the eventual outcome, as suggested by
\citet{1990ApJ...365...86R}. The group's H$\alpha$ luminosity is
$1\times 10^{42}$\,erg\,s$^{-1}$ \citep[][ adjusted to our
distance]{1997ApJ...479..190I}, implying a star formation rate of
$7.9\,M_{\sun}$\,yr$^{-1}$ \citep{1998ARA&A..36..189K}. Although
high, this rate of star formation is sustainable for a long time,
at least another 3\,Gyr given the group's \ion{H}{i} mass of
$2.5\times 10^{10}\,M_{\sun}$ \citep{1991AJ....101.1957W}. Most of
this \ion{H}{i} mass is contained within an area of
$3^{\prime}\times 1^{\prime}$ \citep{1991AJ....101.1957W},
$8.6\times 10^{8}$\,pc$^2$ at our adopted distance, implying a gas
surface mass density of about $30\,M_{\sun}$\,pc$^{-2}$. Supposing
a conversion efficiency from \ion{H}{i} gas into stars of
approximately 10\% at this surface density
\citep{1998ARA&A..36..189K} and that at least 50-60\% of the mass
converted to stars will produce long-lived stars
\citep{1991A&AS...87..109K}, we infer that the eventual outcome
could be a galaxy with a stellar mass of several $10^9\,M_{\sun}$.
This mass is several times larger than the current mass estimates:
the colours from \citet{1997ApJ...479..190I} are completely
dominated by the light from young stars, and therefore imply only
lower limits to the stellar mass of order $3\times
10^8\,M_{\sun}$, based upon the colours of the single stellar
populations from \citet{1994ApJS...94...63B}. Therefore, it is
reasonable to expect an eventual metal production at least equal
to that which has already occurred, accompanied by a similar
increase in the chemical abundances.  Several Gyr after the
cessation of star formation, a galaxy with a stellar mass of
several $10^9\,M_{\sun}$ would have a luminosity of $M_B \approx
-18$\,mag \citep[e.g.,][]{1994ApJS...94...63B} and its oxygen
abundance would place it above the metallicity-luminosity relation
for dwarf irregulars.  Given that the velocity gradients in the
\ion{H}{i} gas found by \citet{1991AJ....101.1957W} imply
generally radial inflow, rather than rotation, we infer that the
probable outcome will be an elliptical galaxy.  Although the
foregoing is entirely speculative, it illustrates that HCG31 could
very plausibly produce a typical low-luminosity elliptical with
normal chemical abundances \citep[e.g.,][]{1998A&A...340...67R}.

\section{Conclusions}

Somewhat at variance with previous studies, it is less evident to
us that HCG31 is a clear example of an on-going merger.  It is
undeniable, however, that there is significant tidal interaction
underway among the members of this compact group. Given their
kinematics, galaxies A, C, and the tidal dwarfs to the north-east
identified by \citet{1996ApJ...462...50H} all appear to be a
single entity at present (the A+C complex). Galaxy E is probably
also an integral part of this complex, otherwise it is a tidal
fragment that has very recently separated. Galaxy F has both
kinematics and chemical abundances that indicate that it is a
tidal fragment that was part of the A+C complex in the relatively
recent past. At this point, galaxies B and G remain kinematically
distinct from the A+C complex.  Finally, we find a velocity for
galaxy Q \citep{1990ApJ...365...86R} that is compatible with it
also being a member of HCG31.

We propose that a gravitational interaction between galaxy G and
the A+C complex is the most complete explanation of the events
occurring in HCG31.  This scenario naturally accounts for the
positions and kinematics of galaxies A, C, E, F, G, and the tidal
dwarf candidates of \citet{1996ApJ...462...50H}, the morphology
and kinematics of the \ion{H}{i} \citep{1991AJ....101.1957W}, the
chemical abundances, and the coordinated star formation in
galaxies separated by up to 40\,kpc
\citep{1997ApJ...479..190I,2000AJ....119.2146J}. In this case, the
A+C complex may have always been a single entity, a late-type
spiral, and its current irregular morphology, high luminosity,
blue colours, and anomalous chemical abundance may simply be a
consequence of multiple bursts of star formation as a result of
strong gas inflows induced by its interaction with galaxy G
\citep{1996ApJ...471..115B,1996ApJ...464..641M}. If this is the
case, a merger may still eventually occur between galaxy G and the
A+C complex given the ample evidence for a strong gravitational
interaction.  While there is no evidence that the kinematical
properties of HCG31 are not the result of an ongoing merger
between galaxies A and C, many additional details are not
accounted for by this solution, particularly the coordinated star
formation and the location of galaxy G.

Based upon its oxygen abundance, its broad-band and H$\alpha$
luminosities, and its kinematics, at present galaxy B appears to
be a relatively normal late-type spiral or bright dwarf irregular
seen nearly edge-on.  Its evolution is probably now being affected
by the group environment, since its colours are very blue and
indicate that its current rate of star formation exceeds its
historical average rate.

\begin{acknowledgements}

MGR thanks Jack Sulentic, H\'ector Aceves, and H\'ector Velazquez
for very helpful discussions.  We also thank the referee, Philippe
Amram, for careful and useful comments.  MGR thanks the staff of
the Observatorio Astrof\'\i sico Guillermo Haro for their
hospitality and Ra\'ul Gonz\'alez and Gerardo Miramon for their
able assistance with the observations. MR and LG thank Gabriel
Garcia and Gustavo Melgoza for their assistance with the
observations at the Observatorio Astron\'omico Nacional in San
Pedro M\'artir. MGR acknowledges financial support from DGAPA
project IN100799 and CONACyT project 37214-E. MR, MGR, and AB
acknowledge financial support from CONACyT project 27984-E and
DGAPA project IN122298. LG acknowledges financial support from
DGAPA project IN113999.  AB acknowledges a Mutis graduate
fellowship from the Agencia Espa\~nola de Cooperaci\'on
Internacional.  MVG acknowledges a CONACyT scholarship (number
114735), and also thanks the time allocation committee at INAOE
for assigning time to this project.

\end{acknowledgements}

%\begin{thebibliography}{}
\bibliographystyle{apj}
\bibliography{h3095b}
%\end{thebibliography}

\end{document}